# The force acting on a polarizable nanoparticle in the quantized electromagnetic field

**Vanik E. Mkrtchian**

*Institute for Physical Research, Armenian National Academy of Sciences, Ashtarak-2, 378410, Republic of Armenia*

In this letter we derive an expression for the force acting on a small (still macroscopic) particle in the field of the quantized electromagnetic radiation in any arbitrary quantum state. This result unifies in one simple formula all known expressions for the forces (i.e. van der Waals or frictional) acting on a small particle.

## 1. Introduction

Growing successes in nanotechnologies and possibility of non-contact control of nanoparticles by optical tweezers [1], [2], or by usage of effects of quantum friction [3] stimulate investigations of the force acting on a polarizable nanoparticle in the external electromagnetic field for the states of the field out of equilibrium. Inclusion in consideration of nonequilibrium states is stipulated by the possibility of tuning the interaction in both strength and sign [4], [5].

The problem of the mechanical force acting on a small, neutral particle, or on an atom with the electromagnetic field in equilibrium is well understood [6]. Meanwhile the same problem for the nonequilibrium states of the field is solved only in a special case when usage of the fluctuation-dissipation theorem provides an opportunity in "construction" of a "hot" half space [4], [5].

In this letter we derive an expression for the force acting on a small (still macroscopic) particle (which we'll call nanoparticle) in the field of the quantized electromagnetic radiation in any arbitrary quantum state. This result provides an opportunity to unify in one expression for the force the all known results related to the problem. As the simplest consequence of the generalization we recover expressions for particle-wall interaction force and expression for the frictional force acting on a moving particle through the blackbody radiation.

## 2. Expression for the force

Compared with the case of atom-field interaction problem the case of macroscopic nanoparticle permits a very important approximation, i.e., in this case for moderate intensities we can ignore dynamics of the particle under influence of the external electromagnetic field and suppose that the particle stays in its initial state (local thermodynamic equilibrium) during all the time of interaction with the external field. Besides, in case of nanoparticles the characteristic size of the particle is much smaller than the wavelength of radiation (for thermal fields relevant wavelengths are of the order $2\pi c/T$ ), so we can solve the problem in Rayleigh regime, [7] i.e., we may ignore the change of external field in the volume of the particle and suppose that the particle behaves as a single electric dipole.

In electric - dipole approximation the force of radiation on a neutral particle located at the point $\vec{r}_A$ is given by [8]

$$\vec{F}(\vec{r}_A,t) = \vec{\nabla}\left\langle \hat{\vec{d}}(t)\hat{\vec{E}}(\vec{r},t)\right\rangle_{\vec{r}=\vec{r}_A}. \tag{1}$$

Taking interaction $\hat{V}$ of electromagnetic field with dipole in the form $\hat{V} = -\hat{\vec{d}}\hat{\vec{E}}(\vec{r}_A)$ we can evaluate $\left\langle \hat{\vec{d}}(t)\hat{\vec{E}}(\vec{r},t)\right\rangle$ in (1) using Keldysh technique [9]:

$$\left\langle \hat{\vec{d}}(t)\hat{\vec{E}}(\vec{r},t)\right\rangle = \left\langle \hat{T}_C \hat{\vec{d}}(t^2)\hat{\vec{E}}(\vec{r},t^1)\hat{S}_C \right\rangle$$

where

$$\hat{S}_C = \hat{T}_C \exp\left[\sum_{\sigma=1,2}(-1)^\sigma i \int_{-\infty}^{+\infty} \hat{V}(\tau^\sigma)d\tau^\sigma \right]$$

Then, in the first order of interaction $\hat{V}$ we find

$$\left\langle \hat{\vec{d}}(t)\hat{\vec{E}}(\vec{r},t)\right\rangle = \sum_{\sigma=1,2}(-1)^{\sigma+1} Tr \int_{-\infty}^{+\infty} d\tau \hat{D}'^{1\sigma}(\vec{r},t;\vec{r}_A,\tau)\hat{\alpha}^{\sigma 2}(\tau,t) \tag{2}$$

where

$$\hat{D}'^{1\sigma}(\vec{r},t;\vec{r}_A,\tau) = \frac{i}{c^2}\partial_t \partial_\tau \hat{D}^{1\sigma}(\vec{r},t;\vec{r}_A,\tau)$$

$\hat{D}^{\lambda\sigma}(\vec{r},t;\vec{r}_A,\tau)$ is the photon propagator

$$\hat{D}_{ij}^{\lambda\sigma}(\vec{r},t;\vec{r}_A,\tau) = -i\left\langle \hat{T}_C \hat{A}_i(\vec{r},t^\lambda)\hat{A}_j(\vec{r}_A,\tau^\sigma)\right\rangle$$

in Dzyaloshinskii gauge ($\varphi = 0$) and

$$\alpha_{ij}^{\lambda\sigma}(t,\tau) = i\left\langle \hat{T}_C \hat{d}_i(t^\lambda)\hat{d}_j(\tau^\sigma)\right\rangle$$

is the particle propagator in the interaction picture.

After Keldysh transformation [9], [10]

$$G^{\lambda\sigma} = \frac{1}{2}\left[G^K + (-1)^{\lambda+1}G^A + (-1)^{\sigma-1}G^R\right]$$

in (2) we come to the expression

$$\left\langle \hat{\vec{d}}(t)\hat{\vec{E}}(\vec{r},t)\right\rangle = \frac{1}{2}Tr\int_{-\infty}^{+\infty}d\tau\left[\hat{D}'^K(\vec{r},t;\vec{r}_A,\tau)\hat{\alpha}^A(\tau,t) + \hat{D}'^R(\vec{r},t;\vec{r}_A,\tau)\hat{\alpha}^K(\tau,t)\right]$$

which for stationary states of electromagnetic field, i.e., when

$$\hat{D}^{K,R}(\vec{r},t;\vec{r}_A,\tau) = \int \frac{d\omega}{2\pi} e^{-i\omega(t-\tau)} \hat{D}^{K,R}(\omega;\vec{r},\vec{r}_A)$$

reduces to a time independent quantity and then for the force (1) we find

$$\vec{F}(\vec{r}_A) = i\int_{-\infty}^{+\infty} \frac{\omega^2 d\omega}{4\pi c^2} Tr\left[\hat{\alpha}^A(\omega)\vec{\nabla}\hat{D}^K(\omega;\vec{r},\vec{r}_A) + \hat{\alpha}^K(\omega)\vec{\nabla}\hat{D}^R(\omega;\vec{r},\vec{r}_A)\right]_{\vec{r}=\vec{r}_A} \quad (3)$$

Further on we'll suppose that the particle stays in a local thermodynamic equilibrium at temperature $T$ during the interaction with electromagnetic field, therefore the Keldysh function of the particle is given by:

$$\hat{\alpha}^K(\omega) = 2i\coth\left(\frac{\omega}{2T}\right) \operatorname{Im}\hat{\alpha}^R(\omega) \quad (4)$$

Expression (3) (with (4)) for the force acting on a small particle in the external electromagnetic field in any arbitrary quantum state is the main result of this paper. This expression also contains all the previous results related to the problem.

### 3. Casimir-Polder interaction with the wall

For instance, in the case of the global equilibrium

$$\hat{D}^K(\omega;\vec{r},\vec{r}_A) = 2i\coth\left(\frac{\omega}{2T}\right)\operatorname{Im}\hat{D}^R(\omega;\vec{r},\vec{r}_A) \quad (5)$$

Expressions (3-4) result in the force acting on the particle in equilibrium [11]:

$$\vec{F}(\vec{r}_A) = \int_{-\infty}^{+\infty} \frac{\omega^2 d\omega}{\pi c^2} \frac{1}{e^{-\omega/T}-1} \operatorname{Im} Tr\left[\hat{\alpha}^R(\omega)\vec{\nabla}\hat{D}^R(\omega;\vec{r},\vec{r}_A)\right]_{\vec{r}=\vec{r}_A} \quad (6)$$

Using analytical properties of retarded functions $\hat{\alpha}^R, \hat{D}^R$, in integrand we can simplify expression (6) representing it as a sum over imaginary frequencies. Really, replacing integration over real frequencies ω by integration in complex upper half plane and then using residue theorem we come to the Matsubara representation of the force in equilibrium [12]:

$$\vec{F}(\vec{r}_A) = T\sum_{s=0}^{+\infty}(2-\delta_{s,0})k_s^2 Tr\left[\hat{\alpha}(i\zeta_s)\vec{\nabla}\hat{D}(i\zeta_s;\vec{r},\vec{r}_A)\right]_{\vec{r}=\vec{r}_A} \quad (7)$$

$$(k_s = \zeta_s/c)$$

As a simple application of this expression let us consider interaction force of the particle with a dielectric half space $z<0$. Insertion of the expression for temperature Green function $\hat{D}(i\zeta_s;\vec{r},\vec{r}_A)$ [13] we find for unique nonzero z component of the force (7) in this case

$$F_z(z_A) = \frac{T}{\pi}\sum_{s=0}^{+\infty}\left(1-\frac{1}{2}\delta_{s,0}\right)\int \vec{k}_\perp d\vec{k}_\perp \left[R-\overline{R}\right]e^{-2w_0 z_A} \quad (8)$$

$$R = \frac{w_0-w}{w_0+w}k_s^2\left[n_y^2\alpha_{xx}(i\zeta_s)+n_x^2\alpha_{yy}(i\zeta_s)\right] \quad (9.a)$$

$$\overline{R} = \frac{w_0\varepsilon(i\zeta_s)-w}{w_0\varepsilon(i\zeta_s)+w}\{k_\perp^2\alpha_{zz}(i\zeta_s)+w^2{}_0\left[n_x^2\alpha_{xx}(i\zeta_s)+n_y^2\alpha_{yy}(i\zeta_s)\right]\} \quad (9.b)$$

Where $w_0 = \sqrt{k_s^2 + k_\perp^2}$, $w = \sqrt{\varepsilon(i\zeta_s)k_s^2 + k_\perp^2}$ and $\vec{n} = \vec{k}_\perp / k_\perp$. For the isotropic particle $\alpha_{ij} = \delta_{ij}\alpha$ expressions (8), (9) are coincident with well-known result [14]

$$F_z(z_A) = T\sum_{s=0}^{+\infty}(2-\delta_{s,0})k_s^4 \alpha(i\zeta_s)\int_0^{+\infty} pdp\, e^{-2k_s p z_A}\left[r + (1-2p^2)\bar{r}\right],$$

where

$$r = \frac{p-s}{p+s},\ \bar{r} = \frac{\varepsilon(i\zeta_s)p - s}{\varepsilon(i\zeta_s)p + s},\ s = \sqrt{\varepsilon(i\zeta_s) - 1 + p^2}.$$

## 4. Frictional force in free space

As an application of our result (3) to the case of frictional forces let us consider the simplest problem: frictional force acting on the moving particle through the blackbody radiation [15]. In the case of free electromagnetic field retarded (advanced) function depends of difference of coordinates [16] and

$$\operatorname{Im}\hat{D}^R(\omega;\vec{k}) = \hat{T}(\vec{k})[\delta(\omega+ck) - \delta(\omega-ck)] \tag{10}$$

$$T_{ij}(\vec{k}) = \frac{2\pi c^2}{k}\left(\delta_{ij} - \frac{k_i k_j}{k^2}\right).$$

In the reference system of the particle moving uniformly with velocity $\vec{v}$ relative to the black body radiation photon distribution function depends also on photon momentum $\vec{k}$ and is given by

$$h(\omega,\vec{k}) = \coth\left(\frac{\omega - \vec{k}\vec{v}}{2T}\right) \tag{11}$$

and for the Keldysh function we may take expression

$$\hat{D}^K(\omega;\vec{k}) = 2ih(\omega,\vec{k})\operatorname{Im}\hat{D}^R(\omega;\vec{k}) \tag{12}$$

In this case we get for (3):

$$F_i = \frac{v}{2\pi c^5 T}\int_0^{+\infty}\frac{\omega^5 d\omega}{\sinh^2(\omega/2T)}f_i(\omega) \tag{13.a}$$

$$f_i(\omega) = \frac{2}{15}\operatorname{Im}\{2\delta_{iz}\operatorname{Tr}\hat{\alpha}(\omega) - [\delta_{ix}\alpha_{xz}(\omega) + \delta_{iy}\alpha_{yz}(\omega) + \delta_{iz}\alpha_{zz}(\omega)]\} \tag{13.b}$$

For isotropic polarization $\alpha_{ij} = \delta_{ij}\alpha$

$$f_i(\omega) = \delta_{iz}\frac{2}{3}\operatorname{Im}\alpha(\omega)$$

and for the force acting on the particle (in the frame of particle) we find [17], [18]

$$\vec{F} = \frac{\vec{v}}{3\pi c^5 T} \int_0^{+\infty} d\omega \frac{\omega^5 \operatorname{Im}\alpha(\omega)}{\sinh^2(\omega/2T)}$$

As we see, expression (3) for the force acting on a nanoparticle in external electromagnetic field in arbitrary quantum state describes in equal foot also Van der Waals interactions of the particle with surroundings and frictional forces acting on the moving particle.

**Acknowledgements**
I'm very grateful to Professors L. Novotny and C. Henkel for stimulating discussions.

[1] L. Novotny, R. Bian, X. Xie, Phys.Rev.Lett.,**79**, 645 (1997).
[2] D. Grier, Nature **424**, 810 (2003).
[3] A.Volokitin, N.Persson, Usp.Fiz.Nauk, **177**, 921 (2007).
[4] C. Henkel, K.Joulain, J.Mulet, J.Greffet, J.Opt.A, **4**, S109 (2002).
[5] M. Antezza, L. Pitaevskii, S. Stringeri, V. Svetovoy, Phys.Rev.A, **77,** 022901 (2008).
[6] S. Buhmann, S. Scheel, Prog.Quant.Elect., **31**, 51(2007)
[7] H. C. van de Hulst, *Light Scattering by Small Particles*, (Dover, New York, NY, 1981).
[8] J. Gordon, A. Ashkin, Phys. Rev. A, **21**, 1606 (1980).
[9] L. Keldysh, Sov.Phys.-JETP, **20**, 1018 (1965).
[10] J. Rammer, H. Smith, Rev.Mod.Phys., **58**, 323 (1986).
[11] L. Novotny, C. Henkel, Opt.Lett., **33**, 1029 (2008).
[12] S. Buhmann, S. Scheel, Phys.Rev.Lett., **100**, 253201(2008).
[13] A. Maradudin, D. Mills, Phys.Rev.B, **11**, 1392 (1975).
[14] V. Parsegian, *Van der Waals Forces*, (Cambridge University Press, New York, 2006).
[15] A. Einstein, Mitt.Phys.,Ges.**18**, 47 (1916).
[16] E. Lifshitz, L. Pitaevskii, *Statistical Physics, Part 2,* (Pergamon, New York, 1981).
[17] V. Mkrtchian, V. Parsegian, R. Podgornik, W. Saslow, Phys.Rev.Lett., **91**, 220801 (2003).
[18] J. Zurita-Sanchez, J. Greffet, L. Novotny, Phys.Rev.A, 69, 022902 (2004).